\documentclass[11pt]{article}
\pdfoutput=1
\usepackage{bbm}
\usepackage[final]{graphics}
\usepackage{graphicx}
\usepackage{amsmath}
\usepackage{amsfonts,amsbsy}
\usepackage{amssymb}
\usepackage{bm}
\usepackage{color}
\usepackage[utf8]{inputenc}

\def\empile#1\over#2{\mathrel{\mathop{\kern 0pt#1}\limits_{#2}}}
\def\bs{\boldsymbol}
\def\wt#1{\widetilde{#1}}

\def\TODO#1{}

\def\p{{\boldsymbol p}}

\def\Q{{\boldsymbol Q}}
\def\E{{\boldsymbol E}}
\def\B{{\boldsymbol B}}

\def\k{{\boldsymbol k}}

\def\x{{\boldsymbol x}}
\def\y{{\boldsymbol y}}

\def\bQbar{\overline{{\boldsymbol Q}}}

\newcommand{\slL}{\raise.15ex\hbox{$/$}\kern-.53em\hbox{$L$}}
\newcommand{\slP}{\raise.15ex\hbox{$/$}\kern-.53em\hbox{$P$}}
\newcommand{\slD}{\raise.15ex\hbox{$/$}\kern-.67em\hbox{$D$}}
\newcommand{\slp}{\raise.1ex\hbox{$/$}\kern-.63em\hbox{$p$}}
\newcommand{\slq}{\raise.1ex\hbox{$/$}\kern-.53em\hbox{$q$}}
\newcommand{\slv}{\raise.1ex\hbox{$/$}\kern-.63em\hbox{$v$}}
\newcommand{\slR}{\raise.15ex\hbox{$/$}\kern-.53em\hbox{$R$}}
\newcommand{\slQ}{\raise.15ex\hbox{$/$}\kern-.53em\hbox{$Q$}}
\newcommand{\slK}{\raise.15ex\hbox{$/$}\kern-.53em\hbox{$K$}}
\newcommand{\slk}{\raise.15ex\hbox{$/$}\kern-.53em\hbox{$k$}}
\newcommand{\slSigma}{\raise.15ex\hbox{$/$}\kern-.53em\hbox{$\Sigma$}}
\newcommand{\slcalP}{\raise.15ex\hbox{$/$}\kern-.63em\hbox{$\cal P$}}
\newcommand{\slcalA}{\raise.15ex\hbox{$/$}\kern-.63em\hbox{$\cal A$}}
\newcommand{\slA}{\raise.15ex\hbox{$/$}\kern-.73em\hbox{$A$}}
\newcommand{\slbfA}{\raise.15ex\hbox{$/$}\kern-.73em\hbox{${\imb A}$}}
\newcommand{\slpartial}{\raise.15ex\hbox{$/$}\kern-.53em\hbox{$\partial$}}
\newcommand{\sla}{\raise.15ex\hbox{$/$}\kern-.53em\hbox{$a$}}
\newcommand{\slb}{\raise.15ex\hbox{$/$}\kern-.53em\hbox{$b$}}
\newcommand{\slc}{\raise.15ex\hbox{$/$}\kern-.53em\hbox{$c$}}
\newcommand{\slC}{\raise.15ex\hbox{$/$}\kern-.63em\hbox{$C$}}
\newcommand{\sln}{\raise.15ex\hbox{$/$}\kern-.575em\hbox{$n$}}

\begin{document}

\date{}

\title{\bf Gauge ambiguity of the quark spectrum\\ in the Color Glass Condensate}

\author{Fran\c cois Gelis${}^{~a}$, Naoto Tanji${}^{~b}$}
\maketitle
 \begin{center}
   \begin{itemize}
  \item[{\bf a.}] Institut de Physique Th\'eorique, Universit\'e Paris-Saclay\\
 CEA, CNRS, F-91191 Gif-sur-Yvette, France
  \item[{\bf b.}] European Centre for Theoretical Studies in Nuclear Physics\\ and Related Areas (ECT*
) and Fondazione Bruno Kessler\\
Villa Tambosi, Strada delle Tabarelle 286, I-38123 Villazzano, Italy
  \end{itemize}
 \end{center}
\vglue 1cm

\begin{abstract}
  In the Color Glass Condensate, the inclusive spectrum of produced
  quarks in a heavy ion collision is obtained as the Fourier transform
  of a $2$-fermion correlation function. Due to its non-locality, the
  two points of this function must be linked by a Wilson line in order
  to have a gauge invariant result, but when the quark spectrum is
  evaluated in a background that has a non-zero chromo-magnetic field,
  this procedure suffers from an ambiguity related to the choice of
  the contour defining the Wilson line. In this paper, we use an
  analytically tractable toy model of the background field in order to
  study this contour dependence. We show that for a straight contour,
  unphysical contributions to the spectrum in $p_\perp^{-2}$ and
  $p_\perp^{-3}$ cancel, leading to a spectrum with a tail in
  $p_\perp^{-4}$. If the contour defining the Wilson line deviates
  from a straight line, the path dependence is at most of order
  $p_\perp^{-5}$ if its curvature is bounded, and of order
  $p_\perp^{-4}$ otherwise. When the contour is forced to go through a
  fixed point, the path dependence is even larger, of order $p_\perp^{-2}$.
\end{abstract}
\vglue 1cm

\section{Introduction}
High energy collisions between hadrons or nuclei probe the
constituents of the projectiles that carry a very small fraction of
the available longitudinal momentum. In this kinematical domain, the
gluon occupation number rises, and saturates at non-perturbatively
large values of order of the inverse strong coupling,
$\alpha_s^{-1}$. The domain where exists such a large gluon occupation
number is delimited by virtualities $Q^2\le Q_s^2(x)$, where $Q_s(x)$
is a momentum scale that depends on the longitudinal momentum fraction
$x$ (and therefore of the collision energy), called the {\sl
  saturation momentum}.

In addition to controlling the onset of gluon saturation, the scale
$Q_s(x)$ is also the typical transverse momentum of the saturated
gluons. This means that it also controls whether multiple scattering
corrections are important: a process whose characteristic transverse
momentum scale is of order $Q_s(x)$ or smaller receives important
corrections due to scatterings on multiple gluons. In particular, this
plays an important role in the question of quark production. In this
case, the characteristic momentum scale is the {\sl transverse mass},
$M_\p^2\equiv (m^2+p_\perp^2)^{1/2}$, that combines the transverse
momentum of the produced quark and its mass. When $M_\p\gg Q_s$, quark
production is perturbative, i.e. dominated by the usual two-gluon
process, $gg\to Q\overline{Q}$. In contrast, when $M_\p\lesssim Q_s$,
processes involving more than two incoming gluons are equally
important.  However, the usual framework of collinear factorization is
not well suited to cope with these multi-gluon corrections, since in
its standard form it provided only a handle on the single-gluon
distribution inside a hadronic projectile. Instead, one can use the
Color Glass Condensate (CGC) effective theory
\cite{Iancu:2002xk,Weigert:2005us,Lappi:2010ek,Gelis:2010nm,Gelis:2012ri}, 
in which a high energy
projectile is described by a color current on the light-cone \cite{McLerran:1993ni,McLerran:1993ka}. At
lowest order, the CGC description is purely classical, thanks to the
large gluon occupation number.

In this framework, the collision of two hadrons or nuclei produces a
color field of order $A^\mu \sim g^{-1}$, determined by solving the
classical Yang-Mills equations \cite{Krasnitz:1998ns,Krasnitz:1999wc,Lappi:2003bi}
with a source made of the superposition of the color currents of the
two projectiles (the boundary conditions should be retarded for
inclusive observables) \cite{Kovner:1995ja,Blaizot:2008yb}.  Quark
production can then be viewed as a problem of particle production in a
time-dependent external field
\cite{Gelis:2004jp,Gelis:2005pb,Hebenstreit:2013qxa,Hebenstreit:2013baa,Kasper:2014uaa,Gelis:2015eua,Gelis:2015kya,Gelfand:2016prm,Tanji:2017xiw}.
In particular, the inclusive quark spectrum can be expressed in terms
of solutions of the Dirac equation in this external field. More
precisely, one needs the {\sl mode functions} of the Dirac equation,
i.e. spinors whose initial condition is a remote past is a free spinor
and that are evolved over the external field up to the time at which
the quark spectrum is evaluated. The quark spectrum is then obtained
as the spatial Fourier transform of an equal-time bilinear combination
of these mode functions, $\psi^\dagger(t,\x)\psi(t,\y)$.

Note however that a non-local object such as
$\psi^\dagger(t,\x)\psi(t,\y)$ is not gauge invariant. This can be
remedied by inserting a Wilson line between the two spinors,
\begin{align*}
  \psi^\dagger(t,\x)\psi(t,\y)
  \to
  \psi^\dagger(t,\x)\,W_\gamma(\x,\y)\,\psi(t,\y),
\end{align*}
defined on a contour $\gamma$ that connects the points $\x$ and
$\y$. When the external color field at the time $t$ where the quark
spectrum is evaluated is null (in fact, any pure gauge works for this
argument), the Wilson line $W_\gamma$ is independent of the path
$\gamma$ and the above modification suffers no ambiguity. But there
are situations where the external field at the time $t$ is not a pure
gauge (this is the case in the CGC description of heavy ion collisions
at a finite time after the collision). When this happens, the above
definition --although gauge invariant-- is ambiguous because it
generally depends on the choice of the path $\gamma$. Note that this
is not an issue in collisions between a dilute and a dense
--saturated-- projectile, since one may expand the yield in powers of
the density of color charges in the dilute projectile. The final state
dressing by a Wilson line would be of higher order in this density
\cite{Gelis:2001da,Kharzeev:2003sk,Blaizot:2004wv,Tuchin:2004rb}.

In this paper, we construct a toy configuration of CGC color currents,
using the $SU(2)$ group for additional simplicity, such that the
classical color fields are simple enough to allow a semi-analytic
discussion of the dependence of the quark spectrum on the path
$\gamma$.  Our paper is organized as follows. In the section
\ref{sec:setup}, we describe the setup of the external field. Then, in
the section \ref{sec:spectrum}, we recall the expression of the quark
spectrum in terms of the fermionic mode functions and evaluate it at a
proper time $\tau=0^+$, i.e. just after the collision. We also discuss
the pathologies of this spectrum when no Wilson line is included to
make it gauge invariant. In the section \ref{sec:linear}, we study the
effect of a Wilson line defined by a straight path between the points
$\x$ and $\y$. We explain how to perform the asymptotic expansion of
the spectrum at large transverse momentum, and derive the leading
term, in order to show that the pathologies encountered without a
Wilson line have now been fixed. The section \ref{sec:curved}
discusses the differences that may arise when using a curved path to
define the Wilson line. We use the non-Abelian Stokes theorem in order
to relate the leading term of the asymptotic expansion of the quark
spectrum to properties of the path $\gamma$, such as its curvature.
Finally, summary and conclusions are in the section \ref{sec:concl},
and a few more technical results are relegated into three appendices.

\section{Background field setup}
\label{sec:setup}
In order to address these questions with a setup that allows an
analytical treatment, we consider an $SU(2)$ gauge group with a very
special configuration of color sources for the two projectiles, which was 
introduced in \cite{Tanji:2018qws}. Let us first recall the link between these sources and the color field
generated immediately after the collision. Starting from the color
current carried by the two projectiles,
\begin{align}
  J^\mu_a(x)\equiv
  \delta^{\mu -}\delta(x^+)\rho_{1a}(\x_\perp)
  +
  \delta^{\mu +}\delta(x^-)\rho_{2a}(\x_\perp), 
\end{align}
we first construct Wilson lines
\begin{align}
  U_m(\x_\perp)\equiv \exp \Big(-ig \frac{1}{{\bs\nabla}_\perp^2} \rho_m(\x_\perp)\Big).
  \label{eq:Umdef}
\end{align}
These Wilson lines enter in the expression of the (light-cone gauge)
color field before the collision for each projectile,
\begin{align}
  \alpha_m^i(\x_\perp)
  =
  \frac{i}{g}\, U_m^\dagger(\x_\perp)\,\partial^i U_m(\x_\perp),\quad \alpha^\pm_m=0.
\end{align}
The non-zero components (we work in the Fock-Schwinger gauge, in which
$A^\tau=0$) of the color field in the forward light-cone just after
the collision (i.e. at a proper time $\tau=0^+$) are then given by \cite{Kovner:1995ja}
\begin{align}
  A^i(\tau=0^+,\x_\perp) = \alpha_1^i(\x_\perp)+\alpha_2^i(\x_\perp),\quad
  A^\eta(\tau=0^+,\x_\perp) = \tfrac{ig}{2}\,\big[\alpha_1^i(\x_\perp),\alpha_2^i(\x_\perp)\big]. 
\end{align}

An analytically tractable setup consists in having homogeneous fields
$\alpha_{1,2}^i$ before the collision, which can be achieved if the
argument of the exponential in $U_{1,2}(\x_\perp)$ is linear,
\begin{align}
U_m(\x_\perp)=\exp\big(i\,\Q_m\cdot \x_\perp\,\sigma^m\big).
\end{align}
To further simplify this setup, the Wilson line $U_1$ has been chosen to
involve only the color generator $\sigma^1$, and $U_2$ involves only
$\sigma^2$. The vectors ${\bs Q}_{1,2}$ are fixed transverse vectors
having the dimension of a mass. With these Wilson lines, we have
\begin{align}
  {\alpha}_m^i =\frac{1}{g}\,Q_m^i \sigma^m, 
\end{align}
and
\begin{align}
  A^i(\tau=0^+)=\frac{1}{g}\,\big(Q_1^i\sigma^1+Q_2^i\sigma^2\big),\quad
  A^\eta(\tau=0^+)=-\frac{1}{g}\,\big(\Q_1\cdot \Q_2\big) \,\sigma^3.
  \label{eq:MVfields}
\end{align}
Since these fields are constant elements of the $SU(2)$ algebra, the
evaluation of Wilson lines in such a background is considerably
simplified.

In order to gain some further insight on the role played by the
orientations of the vectors $\Q_{1,2}$, we can also calculate the
chromo-electric and chromo-magnetic fields just after the collision,
\begin{align}
  \E =\frac{2}{g}\,\big(\Q_1\cdot \Q_2\big)\,\widehat{\bs z}\,\sigma^3,\quad
  \B = \frac{2}{g}\, \big(\Q_1\times \Q_2\big)\,\sigma^3.
  \label{eq:EBfields}
\end{align}
Thus, we see that the radiated field is purely electric if $\Q_1$ and
$\Q_2$ are parallel, and purely magnetic when they are
orthogonal. Intermediate orientations lead to a mixture of electrical
and magnetic fields.

\section{Inclusive quark spectrum}
\label{sec:spectrum}
In the collisions of two hadron/nuclei described by CGC, 
the inclusive spectrum of produced quarks can be expressed as
follows \cite{Gelis:2015eua},
\begin{equation}
\frac{dN}{dy_p d^2 \p_\perp} 
=\frac{1}{8\pi (2\pi )^3 L_\eta } \sum_{s,s^\prime} \sum_{a,a^\prime} \int\! \frac{d^2 \k_\perp}{(2\pi)^2} \frac{d\nu}{2\pi} \frac{d\nu^\prime}{2\pi} \,
\bigg| \left( \widehat\psi_{\p_\perp \nu^\prime s^\prime a^\prime}^{0+} | \widehat\psi_{\k_\perp \nu s a}^{-} \right)_{\tau=0^+} \bigg|^2. 
\label{spectrum}
\end{equation}
In this formula, $\widehat\psi_{\p_\perp \nu^\prime s^\prime a^\prime}^{0+}$
is a free positive energy spinor of transverse momentum $\p_\perp$,
rapidity $\nu'$, spin $s'$ and color $a'$.
$\widehat\psi_{\k_\perp \nu s a}^{-}$ is a spinor that has interacted with
the background color field, starting at $x^0=-\infty$ as a negative
energy spinor of transverse momentum $\k_\perp$, rapidity $\nu$, spin
$s$ and color $a$. The inner product $\big(\cdot\big|\cdot\big)$ in
this formula is defined by
\begin{equation}
\left( \widehat\psi_1 \big| \widehat\psi_2 \right)_\tau = \int d^2 \x_\perp d\eta \; \widehat{\psi}_1^\dagger (\tau ,\x_\perp ,\eta ) \widehat{\psi}_2 (\tau ,\x_\perp ,\eta ).
\end{equation}
Because the background field is invariant under boosts in the
direction of the collision, the squared inner product in
eq.~(\ref{spectrum}) produces two powers of $2\pi
\delta(\nu-\nu')$. One of them is absorbed by the integral over
$d\nu$, while the second power becomes a factor
$2\pi \delta(0)\equiv L_\eta$, that precisely cancels the prefactor
$L_\eta^{-1}$ (unsurprisingly, since we are calculating the spectrum
per unit of rapidity, the length of the rapidity interval under
consideration must cancel from the result).

At very small proper time, the explicit form of the spinors involved
in eq.~(\ref{spectrum}) was derived in \cite{Gelis:2015eua}. They read
\begin{align}
\widehat{\psi}_{\p_\perp \nu s a}^{0+} (x)
&= e^{-\pi i/4} \sqrt{\frac{2}{M_\p}} e^{i(\p_\perp \cdot \x_\perp +\nu \eta)} 
\left\{ e^{\pi \nu/2} \left( \frac{M_\p \tau}{2} \right)^{-i\nu} \Gamma (\tfrac{1}{2}+i\nu) \mathcal{P}^+ \right. \notag \\
&\hspace{10pt} \left. 
+e^{-\pi \nu/2} \left( \frac{M_\p \tau}{2}\right)^{i\nu} \Gamma (\tfrac{1}{2}-i\nu) \mathcal{P}^- \right\}
u_s (\p_\perp ,y=0) \chi_a
\label{psi_free}
\end{align}
and
\begin{align}
\widehat{\psi}_{\k_\perp \nu s a}^- (x)
  &= -\frac{e^{\pi i/4}}{\sqrt{M_\k}}
    \int \frac{d^2 {\bs\ell}_\perp}{(2\pi)^2} \frac{ e^{i({\bs\ell}_\perp \cdot \x_\perp +\nu \eta)} }{M_{\bs\ell} }\nonumber\\
  &  \times\left\{ e^{\pi \nu/2} \left( \frac{M_{\bs\ell}^2 \tau}{2M_\k}\right)^{i\nu} \Gamma (\tfrac{1}{2}-i\nu) U_2^\dagger (\x_\perp ) \widetilde{U}_2 ({\bs\ell}_\perp +\k_\perp ) \gamma^+ \right. \notag \\
&\qquad \left.
+e^{-\pi \nu/2} \left( \frac{M_{\bs\ell}^2 \tau}{2M_\k}\right)^{-i\nu} \Gamma (\tfrac{1}{2}+i\nu) U_1^\dagger (\x_\perp ) \widetilde{U}_1 ({\bs\ell}_\perp +\k_\perp ) \gamma^- \right\}\notag\\
  &\qquad\qquad\qquad\times
    (\ell^i \gamma^i +m) v_s (\k_\perp ,y=0) \chi_a, 
\label{psi-}
\end{align}
where $\chi_a$ ($a=1,2,3$) is a unit vector in color space,
${\cal P}^+ \equiv (\gamma^- \gamma^+)/2 $ and
${\cal P}^- \equiv (\gamma^+ \gamma^-)/2 $ are projectors acting on
the Dirac indices, $M_\k\equiv \sqrt{\k_\perp^2+m^2}$ and
$\wt{U}_m(\p_\perp)$ is the Fourier transform of the Wilson line
defined in eq.~(\ref{eq:Umdef}),
\begin{equation}
\widetilde{U}_m ({\bs\ell}_\perp )
= \int d^2 \x_\perp \; e^{-i{\bs\ell}_\perp \cdot \x_\perp} U_m (\x_\perp ) .
\end{equation}

Eq.~(\ref{spectrum}) is valid only if there is no background
color field present at the time where the quark spectrum is evaluated.
In particular, this formula is not gauge invariant because the
expression (\ref{psi_free}) of the free spinor must be modified if
there is a pure gauge (i.e. gauge equivalent to the null color field)
background field. Let us ignore this difficulty for a brief moment, in
order to see the type of pathology that one would encounter by using
eq.~(\ref{spectrum}) unmodified. For the setup of background field
introduced in the previous section, we have
\begin{align}
U_m (\x_\perp )
= \frac{1+\sigma^m}{2} e^{i\Q_m \cdot \x_\perp} +\frac{1-\sigma^m}{2} e^{-i\Q_m \cdot \x_\perp}
\end{align}
and the Fourier transform reads
\begin{equation}
\widetilde{U}_m ({\bs\ell}_\perp )
= \frac{1+\sigma^m}{2} (2\pi)^2 \delta^2 ({\bs\ell}_\perp -\Q_m )
+\frac{1-\sigma^m}{2} (2\pi)^2 \delta^2 ({\bs\ell}_\perp +\Q_m ).
\end{equation}
Plugging this in the dressed spinor (\ref{psi-}) gives
\begin{align}
\widehat{\psi}_{\k_\perp \nu s a}^- (x)
&= \frac{e^{\pi i/4}}{\sqrt{M_\k}} e^{i(\nu\eta-\k_\perp \cdot \x_\perp)} \notag \\
&\times \bigg\{ e^{\pi \nu/2} \Gamma (\tfrac{1}{2}-i\nu) 
\left[ \tfrac{1}{M_{\k -\Q_2}} \left( \tfrac{M_{\k -\Q_2}^2 \tau}{2M_\k}\right)^{i\nu} \gamma^+ \left( M_\k \gamma^0 -Q_2^i \gamma^i \right) \frac{1+\sigma^2}{2} \right. \notag \\
&\hspace{90pt} \left.
+\tfrac{1}{M_{\k +\Q_2}} \left( \tfrac{M_{\k +\Q_2}^2 \tau}{2M_\k}\right)^{i\nu}  \gamma^+ \left( M_\k \gamma^0 +Q_2^i \gamma^i \right) \frac{1-\sigma^2}{2} \right] \notag \\
&+e^{-\pi \nu/2} \Gamma (\tfrac{1}{2}+i\nu) 
\left[ \tfrac{1}{M_{\k -\Q_1}} \left( \tfrac{M_{\k -\Q_1}^2 \tau}{2M_\k}\right)^{-i\nu} \gamma^- \left( M_\k \gamma^0 -Q_1^i \gamma^i \right) \frac{1+\sigma^1}{2} \right. \notag \\
&\hspace{90pt} \left.
                                                                                                                                                                                             +\tfrac{1}{M_{\k +\Q_1}} \left( \tfrac{M_{\k +\Q_1}^2 \tau}{2M_\k}\right)^{-i\nu}  \gamma^- \left( M_\k \gamma^0 +Q_1^i \gamma^i \right) \frac{1-\sigma^1}{2} \right] \bigg\}\notag\\
  &\hspace{150pt} \times v_s (\k_\perp,y=0 ) \chi_a \, .
\label{psi_uni}
\end{align}
(In the derivation, we have used
$(M_\k \gamma^0 -k^i \gamma^i +m)v_s (\k_\perp ,y=0) =0$.) It is
then straightforward to calculate the inner product between this
spinor and the vacuum spinor (\ref{psi_free}) in order to obtain the
following expression of the quark spectrum,
\begin{align}
\frac{dN}{dy_p d^2 \p_\perp} 
&= \frac{S_\perp}{8\pi^4} \bigg\{ 4 -\sum_{\epsilon_1,\epsilon_2=\pm}
\frac{F\left(\tfrac{M_{\p +\epsilon_1 \Q_1} M_{\p +\epsilon_2 \Q_2}}{M_\p^2} \right)}{M_{\p +\epsilon_1 \Q_1} M_{\p +\epsilon_2 \Q_2}} \notag \\
&\times
\left[ M_\p^2-\epsilon_1 \epsilon_2 \Q_1 \cdot \Q_2 +\epsilon_1 \p_\perp \cdot \Q_1 +\epsilon_2 \p_\perp \cdot \Q_2 +2\epsilon_1 \epsilon_2 \tfrac{(\p_\perp \cdot \Q_1)(\p_\perp \cdot \Q_2)}{M_\p^2} \right] 
\bigg\} , \label{spec_wol}
\end{align}
with $F(x)\equiv\ln( x)/\sinh( \ln x)$ and $S_\perp$ the transverse
section of the overlap region of the colliding nuclei (this comes from
a factor $(2\pi)^2\delta(\p_\perp=0)$). If we expand this expression
at large transverse momentum ($p_\perp\gg Q_{1,2}$), the leading term is of the form
\begin{align}
  \frac{dN}{dy_p d^2 \p_\perp} 
  \sim S_\perp \left\{
  \frac{\Q_1^2}{p_\perp^2}
  \oplus
  \frac{\Q_2^2}{p_\perp^2}
  \right\}.
\end{align}
\begin{figure}[tb]
  \centering
  \includegraphics[width=10cm]{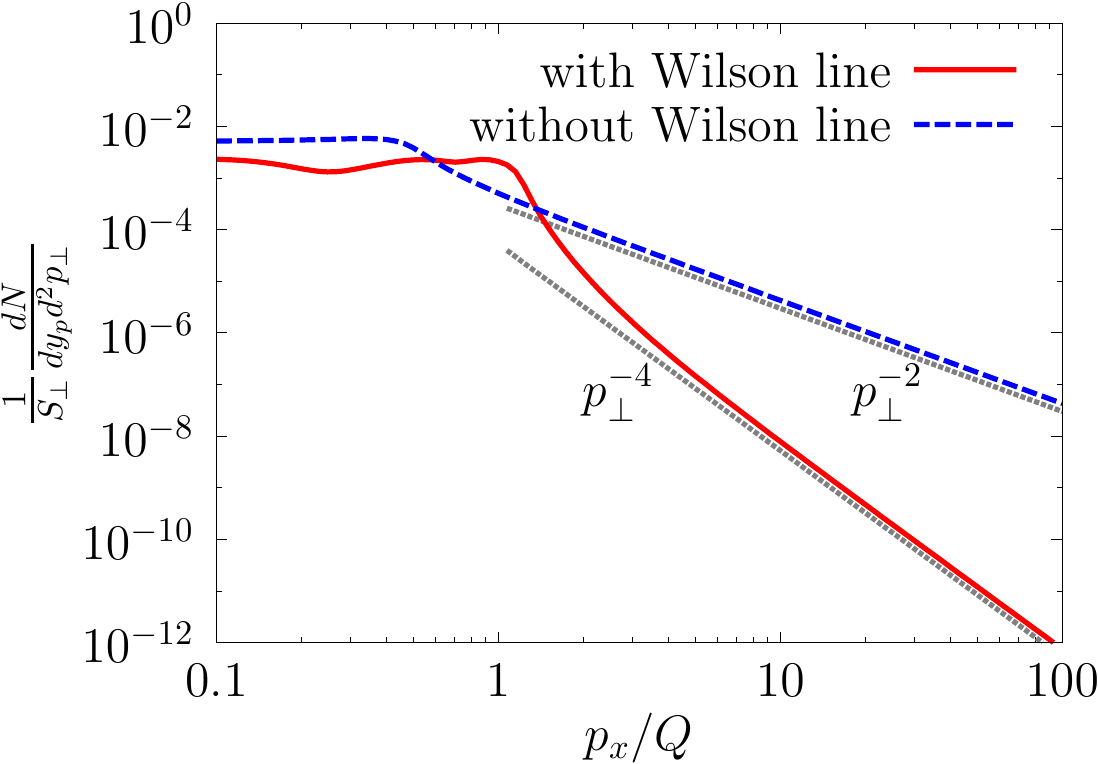}
  \caption{\label{fig:elec}Quark spectrum (for $p_y=0$, plotted as a
    function of $p_x$) in the purely electrical case ($\Q_1$ and
    $\Q_2$ parallel), for $Q_1=Q_2=Q_s/2$ and $m/Q_s=0.1$. Dashed
    blue line: spectrum defined without a Wilson line. Solid red line:
    spectrum defined with a Wilson line. Dotted lines: power laws
    $p_\perp^{-2}$ and $p_\perp^{-4}$.}
\end{figure}
The spectrum \eqref{spec_wol} is plotted in figure \ref{fig:elec} by 
the dashed line. Such a result suffers from two pathologies:
\begin{itemize}
\item Although it vanishes if $\Q_1=\Q_2=0$, it does not vanish if
  only one of the vectors $\Q_{1,2}$ is zero. Physically, this means
  that this formula fails to give a null spectrum when one of the
  projectiles carries no color charge.
\item The power law $p_\perp^{-2}$ is inconsistent with expectations
  from perturbation theory (i.e. an order-by-order expansion in the
  color charge densities of the projectiles -- see for instance
  \cite{Gelis:2003vh,Blaizot:2004wv,Fujii:2006ab}), that indicate that
  the spectrum should decrease as the fourth power of transverse
  momentum, $p_\perp^{-4}$.
\end{itemize}
The two problems are in fact related by a simple dimensional
argument. Indeed, if we take for granted the proportionality of the
spectrum to the transverse area $S_\perp$ (a trivial consequence of
the inclusive nature of the spectrum), the only way to increase the
order in $\Q_{1,2}$ in the numerator (in order to have a term in
$\Q_1^2\Q_2^2$) is to also increase by two units the power in the
denominator.

\section{Effect of a straight Wilson line}
\label{sec:linear}
\subsection{Insertion of a Wilson line}
As already mentioned, the main problem with the formula
(\ref{spectrum}), regardless of the precise background field for which
it is used, is that it assumes that the free positive energy spinors
$\widehat\psi^{0+}$ are the vacuum ones. In particular, this implies a
trivial color structure for the bilinear form constructed with this
spinor,
\begin{align}
  \sum_{a'} \widehat{\psi}_{\p_\perp \nu' s' a'}^{0+} (x)
  \Big[\widehat{\psi}_{\p_\perp \nu' s' a'}^{0+} (y)\Big]^\dagger
  \propto \sum_{a'} \chi_{a'}\chi^\dagger_{a'}={\bs 1}_{\rm color}. 
\end{align}
If we perform a gauge rotation of all the color fields and spinors in
the problem, the spectrum of produced quarks (summed over colors)
should remain unchanged, but this can only be true if we replace
eq.~(\ref{psi_free}) by an expression that depends on the gauge
rotated background field. As explained in \cite{Gelis:2015eua}, when
the background field is a pure gauge, this dressing is achieved in a
non-ambiguous manner by inserting a Wilson line connecting the two
spinors. This amounts to the following replacement,
\begin{align}
  \sum_{a'} \chi_{a'}\chi^\dagger_{a'}
  \quad\to\quad
  W_\gamma(x,y),
  \label{eq:W-subst}
\end{align}
where $W_\gamma(x,y)$ is a Wilson line between the points $x$ and $y$,
defined along a path $\gamma$. Note that this Wilson line can be taken
to be along a purely spatial path since the quark spectrum involves a
pair of spinors evaluated at equal times. Furthermore, since
$A_\eta=-\tau^2 A^\eta=0$ at $\tau=0^+$, the measure $dx^\mu A_\mu$
does not receive any contribution from the rapidity direction (this
statement is only true at the time $\tau=0^+$, that we consider here),
and $W_\gamma$ depends only on the projection of the path $\gamma$ on
the transverse plane.

When the background field at the time where the spectrum is evaluated
is a pure gauge, such a Wilson line depends only on its endpoints, and
the resulting spectrum is therefore non-ambiguous in the sense that it
does not depend on the specific choice of the path $\gamma$. However,
this is not the case in the MV model at a finite proper time, and in
particular with the fields of eqs.~(\ref{eq:MVfields}) (that these
fields are not a pure gauge is a trivial consequence of the fact that
the $\E,\B$ fields are not both zero -- see eqs.~(\ref{eq:EBfields})).
Therefore, we expect a path dependence of the quark spectrum when
evaluated after the substitution of eq.~(\ref{eq:W-subst}).  In order
to study this path dependence, one may first rewrite the quark
spectrum in terms of the Fourier transform of the Wilson line
$W_\gamma$ as follows,
\begin{align}
\frac{dN}{dy_p d^2 \p_\perp} 
&= \frac{2}{(2\pi )^4} \int\! \frac{d^2 \k_\perp}{(2\pi)^2} 
\;\bigg\{
2\,\text{tr} [\widetilde{W}_\gamma (\k_\perp )] \notag \\
&\hspace{10pt} -\frac{1}{2} \sum_{\epsilon_1 ,\epsilon_2 =\pm}
                                                             G(\p_\perp ,\k_\perp ,\epsilon_1 \Q_1 ,\epsilon_2 \Q_2) \, \text{tr} \left[(1+\epsilon_1 \sigma^1 +\epsilon_2 \sigma^2 )\widetilde{W}_\gamma (\k_\perp )\right] \bigg\},
                                                             \label{eq:spectrum-W}
\end{align}
where
\begin{align}
G(\p_\perp ,\k_\perp ,\Q_1 ,\Q_2)
&\equiv \frac{F \left( \frac{M_{\k -\p  -\Q_1} M_{\k-\p  -\Q_2}}{M_\p M_{\k -\p}} \right)}{M_\p M_{\k -\p } M_{\k -\p -\Q_1} M_{\k -\p -\Q_2}} 
 \notag \\
&\hspace{10pt} \times
\bigg[ \left( M_{\k -\p}^2 -(\k_\perp -\p_\perp) \cdot \k_\perp \right) \left( M_{\k -\p}^2 -\Q_1 \cdot \Q_2 \right) \notag \\
&\hspace{30pt} 
+(\p_\perp \cdot \Q_1 ) \left( M_{\k -\p}^2 -(\k_\perp -\p_\perp ) \cdot \Q_2 \right) \notag \\
&\hspace{30pt} 
+(\p_\perp \cdot \Q_2 ) \left( M_{\k -\p}^2 -(\k_\perp -\p_\perp ) \cdot \Q_1 \right) \bigg].
\end{align}
($F(x)$ is the same function as in eq.~(\ref{spec_wol}).) and
\begin{align}
  \wt{W}_\gamma(\k_\perp)\equiv \int d^2\x_\perp d^2\y_\perp\;
  e^{i\k_\perp\cdot (\x_\perp-\y_\perp)}\, W_\gamma(\x_\perp,\y_\perp).
  \label{eq:FTW}
\end{align}
Note that when we do not insert the Wilson line $W_\gamma$ (i.e.,
$W_\gamma\equiv {\bs 1}_{\rm color}$ and
$\wt{W}_\gamma(\k_\perp)=S_\perp (2\pi)^2\delta(\k_\perp)\times{\bs
  1}_{\rm color}$), we recover the expression (\ref{spec_wol}). 
The momentum integral in \eqref{eq:spectrum-W} can be performed analytically 
in the case of a purely electric background ($\Q_1$ parallel to $\Q_2$) as
explained in Appendix~\ref{app:elespec}.
For a purely
electrical background, we do not need to specify the shape of the path
used to defined the Wilson line, so this question can be postponed
until later. The result is shown by the solid red line in the figure
\ref{fig:elec}, where we can see that the tail of the quark spectrum
now decreases as $p_\perp^{-4}$. In the rest of this section, we
consider a straight path $\gamma$, and provide an analytical
explanation for this improved high-$p_\perp$ behavior for general configurations of
the vectors $\Q_{1,2}$. 

\subsection{Asymptotic expansion of the spectrum}
In order to check whether the Wilson line $W_\gamma$ fixes the
incorrect large $p_\perp$ behavior of the spectrum, we need a
systematic way of obtaining the asymptotic expansion of
eq.~(\ref{eq:spectrum-W}). Intuitively, large transverse momenta
$p_\perp$ should correspond to a small separation $\x_\perp-\y_\perp$,
and we thus expect that this asymptotic expansion can be written in
terms of the coefficients of the expansion of the Wilson line in
powers of $\x_\perp-\y_\perp$. Firstly, we write
\begin{equation}
G(\p_\perp ,\k_\perp ,\Q_1 ,\Q_2)
= 1 -\sum_{n\ge 2} \frac{1}{p_\perp^n} G^{(n)} (\widehat\p_\perp ,\k_\perp ,\Q_1 ,\Q_2),
\end{equation}
where $\widehat\p_\perp\equiv \p_\perp/p_\perp$.  (Note that the
expansion of the function $G$ has no term of order $p_\perp^{-1}$.)
The zeroth order term cancels trivially with the term in
$2\,\text{tr} [\widetilde{W}_\gamma (\k_\perp )]$ in
eq.~(\ref{eq:spectrum-W}). The remaining terms in the quark spectrum
can be written as
\begin{align}
\frac{dN}{dy_p d^2 \p_\perp} 
&= \frac{1}{(2\pi )^4} \sum_{n\ge 2} \frac{1}{p_\perp^n} \sum_{\epsilon_1 ,\epsilon_2 =\pm} \notag \\
&\hspace{10pt} \times \int\! \frac{d^2 \k_\perp}{(2\pi)^2} \;
\text{tr} \left[ G^{(n)} (\widehat\p_\perp ,\k_\perp ,\epsilon_1 \Q_1 ,\epsilon_2 \Q_2) \, 
(1+\epsilon_1 \sigma^1 +\epsilon_2 \sigma^2 )\widetilde{W}_\gamma (\k_\perp )\right].
\label{spec_exp}
\end{align}
The explicit form of the coefficient functions $G^{(n)}$ that appear
in this expansion are listed in the appendix \ref{app:coeffs}. These
coefficients are polynomials in the components of the momentum
$\k_\perp$, and therefore we need moments of the form
\begin{align}
\int \frac{d^2\k_\perp}{(2\pi)^2}\;k_\perp^{i_1}\cdots k_\perp^{i_n}\;\wt{W}_\gamma(\k_\perp).
\end{align}
This is where a dependence on the choice of the path $\gamma$ may
arise. Note however that the zeroth order moment is universal, since
\begin{align}
\int \frac{d^2\k_\perp}{(2\pi)^2}\;\wt{W}_\gamma(\k_\perp)=S_\perp\times{\bs 1}_{\rm color}
\end{align}
regardless of the path $\gamma$ (this result is also true if we do not
insert a Wilson line to connect the two spinors). If there is no
Wilson line, all higher moments are identically zero, which gives
the leading term in $p_\perp^{-2}$ obtained in the previous section.

\subsection{Leading term with a straight Wilson line}
These moments are easiest to calculate when we choose a straight
line to connect the points $x$ and $y$. This path will provide a
reference result to compare with when we consider curved paths in the
next section. Let us denote $W_{_L}$ this straight Wilson line. Since
the background color fields are homogeneous with our setup, this
Wilson line has a trivial expression in coordinate space
\begin{align}
  W_{_L}(\x_\perp ,\y_\perp)=\exp\big(-ig{\bs A}_\perp\cdot(\x_\perp-\y_\perp)\big),
\end{align}
with $g{\bs A}_\perp\equiv \Q_1 \sigma^1+\Q_2 \sigma^2$. The first moment of $\wt{W}_{_L}$ reads
\begin{align}
  \int \frac{d^2\k_\perp}{(2\pi)^2}\;k_\perp^i\,\wt{W}_{_L}(\k_\perp)
  &=\int \frac{d^2\k_\perp}{(2\pi)^2}\;k_\perp^i\,
    \int d^2\x_\perp d^2\y_\perp\;
    e^{i\k_\perp\cdot (\x_\perp-\y_\perp)}\, W_\gamma(\x_\perp,\y_\perp)\notag\\
  &=
    S_\perp
    \int \frac{d^2\k_\perp}{(2\pi)^2}\;k_\perp^i\,
    \int d^2\x_\perp \;
    e^{i\k_\perp\cdot \x_\perp}\, W_\gamma(\x_\perp,0)\notag\\
  &=S_\perp gA_\perp^i. 
\end{align}
For moments of order two and higher, the calculation is the same, but
we have to pay attention to the non-commutative nature of
${\bs A}_\perp$. A careful analysis shows that one gets symmetrized
products of the ${\bs A}_\perp$'s,
\begin{align}
  \int \frac{d^2\k_\perp}{(2\pi)^2}\;k_\perp^{i_1}\cdots k_\perp^{i_n}\;\wt{W}_{_L}(\k_\perp)
  =\frac{S_\perp}{n!} \,\Big\{(gA_\perp^{i_1})\cdots(gA_\perp^{i_n})
  +{\rm permutations}\Big\}.
\end{align}
These products are then weighted by a factor
$(1+\epsilon_1 \sigma^1+\epsilon_2\sigma^2)$ and a color trace is
performed. This gives the following replacement rules
\begin{align}
  1&&\to&& 2\notag\\
  k^i_\perp&&\to&& 2(\epsilon_1 Q_1^i+\epsilon_2 Q_2^i)\notag\\
  k^i_\perp k^j_\perp    &&\to&&2(Q_1^iQ_1^j+Q_2^iQ_2^j)\notag\\
  k^i_\perp k^j_\perp k^l_\perp &&\to&& \frac{2}{3!}\Big\{
                                        (Q_1^iQ_1^j+Q_2^iQ_2^j)(\epsilon_1 Q_1^l+\epsilon_2 Q_2^l)+{\rm perms.}
                                        \Big\}\notag\\
  k^i_\perp k^j_\perp k^l_\perp k^m_\perp &&\to&& \frac{2}{4!}
                                                  \Big\{
                                                  (Q_1^iQ_1^j+Q_2^iQ_2^j)(Q_1^lQ_1^m+Q_2^lQ_2^m)+{\rm perms.}
                                                  \Big\}
\end{align}
For the terms in $p_\perp^{-2}$, this leads to a coefficient of the form
\begin{align}
&\int\frac{d^2 \k_\perp}{(2\pi)^2} \;
\text{tr} \left[ G^{(2)} (\widehat\p_\perp ,\k_\perp ,\epsilon_1 \Q_1 ,\epsilon_2 \Q_2) \, 
                (1+\epsilon_1 \sigma^1 +\epsilon_2 \sigma^2 )\widetilde{W}_\gamma (\k_\perp )\right]\notag\\
  &\qquad
  =
  -2\epsilon_1 \epsilon_2 S_\perp \Big(\Q_1\cdot\Q_2-\tfrac{2}{3}(\widehat\p_\perp\cdot\Q_1)(\widehat\p_\perp\cdot\Q_2)\Big),
\end{align}
which gives zero after summation over $\epsilon_{1,2}=\pm$.  Note that
this cancellation results from a delicate interplay between constant
terms, terms in $k_\perp^i$, and terms in $k_\perp^i k_\perp^j$. At
the order $p_\perp^{-3}$, even though the expressions are lengthier
(see the appendix \ref{app:coeffs}), it turns out that the
cancellation happens order by order in $k_\perp^i$, without the need
of combining terms of various orders. The first non-canceling
contributions arise at order $p_\perp^{-4}$ (we have used {\sc
  Mathematica} to extract them, and do not provide here the explicit
form of the coefficient $G^{(4)}$, which is rather unilluminating). At
this order, the quark spectrum reads
\begin{align}
  \frac{dN}{dy_p d^2\p_\perp}
  &=
    \frac{S_\perp}{60\pi^4 p_\perp^4}\,\Big[
    27\, \Q_1^2\Q_2^2
    -12\,(\Q_1\cdot\Q_2)^2
    -8\,(\widehat\p_\perp\cdot\Q_1)^2(\widehat\p_\perp\cdot\Q_2)^2\notag\\
  &\qquad\qquad\quad
    +4\,(\Q_1\cdot\Q_2) (\widehat\p_\perp\cdot\Q_1)(\widehat\p_\perp\cdot\Q_2)\notag\\
  &\qquad\qquad\quad
    -2\,(\widehat\p_\perp\cdot\Q_1)^2\Q_2^2
    -2\,(\widehat\p_\perp\cdot\Q_2)^2\Q_1^2
    \Big]+{\cal O}(p_\perp^{-5}).
    \label{eq:4th-exp}
\end{align}
One may easily check that the coefficient of the term in
$p_\perp^{-4}$ is positive definite, for all orientations of the
vectors $\widehat\p_\perp, \Q_1$ and $\Q_2$. Therefore, the
conclusion of this section is that the insertion of a straight Wilson
line removes the unphysical $p_\perp^{-2}$ and $p_\perp^{-3}$ terms
from the quark spectrum.

\section{Curved Wilson lines}
\label{sec:curved}
\subsection{Non-Abelian Stokes theorem}
\begin{figure}[btp]
  \centering
  \includegraphics[width=60mm]{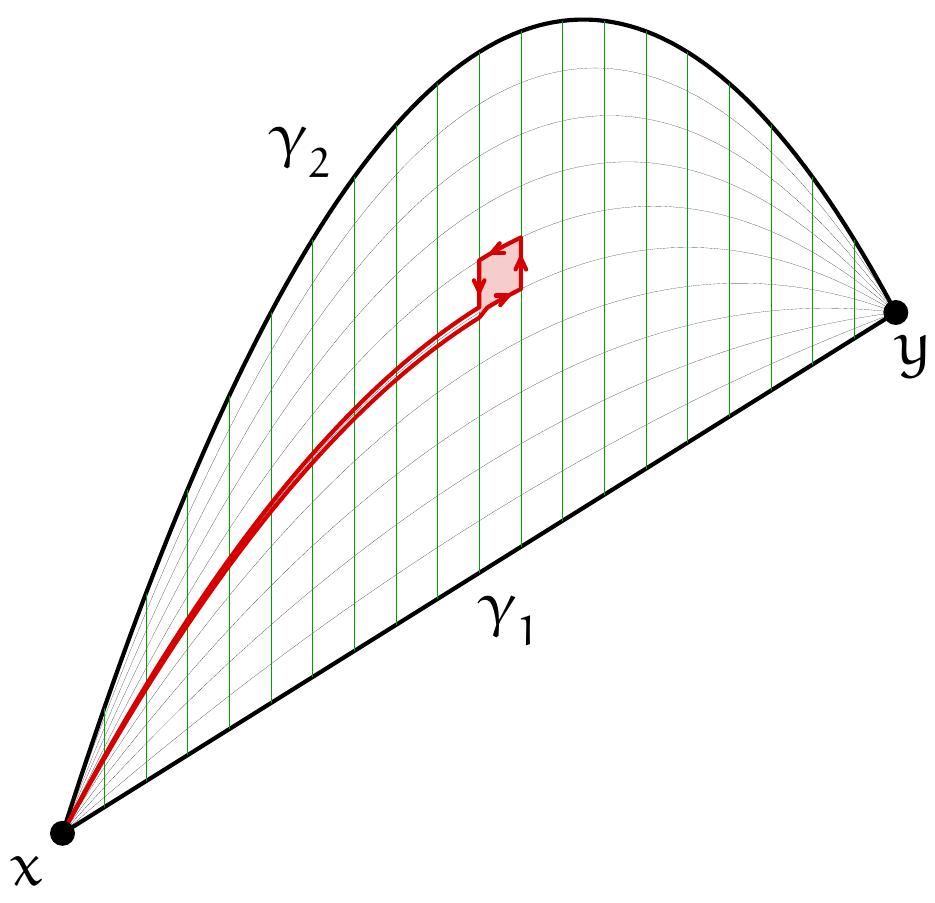}
  \caption{\label{fig:stokes}Illustration of the non-Abelian Stokes
    theorem. The argument of the exponential in
    eq.~(\ref{eq:nab-stokes}) is a sum of terms such as the one
    highlighted in red.}
\end{figure}

After having observed that the insertion of a straight Wilson line in
the quark spectrum leads to an asymptotic behavior in agreement with
perturbative expectations, the obvious question is that of the
universality of this result:
\begin{itemize}
\item does the cancellation of the terms in $p_\perp^{-2}$ and
  $p_\perp^{-3}$ rely on taking a straight path between the two
  points?

\item even if this cancellation works for any path $\gamma$, does the
  leading $p_\perp^{-4}$ depend on the shape of the path?

\item more generally, what is the lowest order at which a path dependence appears?
\end{itemize}
In order to address these questions, we need to compare the effect of
Wilson lines defined on two different contours $\gamma_1$ and
$\gamma_2$. This can be done by considering the quantity
$W_{\gamma_1}W^{-1}_{\gamma_2}$, which is a Wilson loop going from $x$
to $y$ along $\gamma_1$ and returning to $x$ via $\gamma_2$. In the
Abelian case, this Wilson loop would be easily expressible in terms of
the flux of the magnetic field through the loop, thanks to Stokes
theorem. In a non-Abelian theory, the situation is more complicated
because the magnetic field is a non-commuting object, but there is a
formal ``non-Abelian Stokes theorem'' that can be expressed as follows
\cite{Bralic:1980ra}. First, we define a family of curves
$\gamma^\mu(t,s)$ that interpolate between $\gamma_1$ and $\gamma_2$,
\begin{align}
  &\gamma^\mu(t,0)=\gamma^\mu_1(0)=\gamma^\mu_2(0)=x^\mu,\notag\\
  &\gamma^\mu(t,1)=\gamma^\mu_1(1)=\gamma^\mu_2(1)=y^\mu,\notag\\
  &\gamma^\mu(0,s)=\gamma^\mu_1(s),\quad\gamma^\mu(1,s)=\gamma^\mu_2(s).
\end{align}
In words, $t$ is a parameter that labels the curves interpolating
between $\gamma_1$ and $\gamma_2$, while $s$ is a coordinate that
labels the points along each of these curves.  The non-Abelian Stokes
theorem reads (see the figure \ref{fig:stokes} for an illustration)
\begin{align}
  W_{\gamma_1}W^{-1}_{\gamma_2}
  =
  {\rm P}_t \exp \Big\{ig\int_0^1 dt ds\; \frac{\partial \gamma^\mu}{\partial t }\frac{\partial \gamma^\nu}{\partial s}
  W_{\gamma(t,\cdot)}(0,s){F}_{\mu\nu}(\gamma(s,t)) W_{\gamma(t,\cdot)}^{-1}(0,s)
  \Big\},
  \label{eq:nab-stokes}
\end{align}
where $W_{\gamma(t,\cdot)}(0,s)$ denotes the Wilson that goes from
$x^\mu$ to $\gamma^\mu(s,t)$ along the path
$\gamma^\mu(t,\cdot)$. Note that the ordering of the exponential in
this formula applies only to the $t$ integral (the ordering in $s$ is
fully specified by the Wilson lines $W_{\gamma(t,\cdot)}(0,s)$).
Under a gauge transformation $\Omega$, the combination $W_{\gamma(t,\cdot)}(0,s){F}_{\mu\nu}(\gamma(s,t)) W_{\gamma(t,\cdot)}^{-1}(0,s)$ transforms as
\begin{align}
  W_{\gamma(t,\cdot)}(0,s){F}_{\mu\nu}(\gamma(s,t)) W_{\gamma(t,\cdot)}^{-1}(0,s)
  \;\to\;
  \Omega(x)\,
  W_{\gamma(t,\cdot)}(0,s){F}_{\mu\nu}(\gamma(s,t)) W_{\gamma(t,\cdot)}^{-1}(0,s)
  \Omega^\dagger(x),
\end{align}
which is precisely the expected transformation of the Wilson loop
$W_{\gamma_1}W^{-1}_{\gamma_2}$. We see the crucial role played by the
Wilson lines $W_{\gamma(t,\cdot)}(0,s)$ in obtaining the correct
transformation law. We can also foresee that these Wilson lines lead
to a crucial difference between the Abelian and non-Abelian cases
regarding the behavior of a small Wilson loop. In the Abelian case, a
small Wilson loop depends only on the magnetic flux through the loop,
i.e. on the local magnetic field and on the area of the loop. In the
non-Abelian case, the presence of $W_{\gamma(t,\cdot)}(0,s)$ implies
also a dependence upon the length of the paths connecting $x$ to other
points inside the loop.

In the case of interest here, where the paths $\gamma_{1,2}$ are
embedded in the transverse plane, the only component of the field
strength that enters in this formula is $F_{12}$, i.e. the component
of the chromo-magnetic field in the $z$ direction. A trivial
consequence of this observation is that the paths $\gamma_1$ and
$\gamma_2$ give the same Wilson line in the case where the background
field is purely electrical (i.e. when $\Q_{1,2}$ are parallel). Thus,
in this case, the quark spectrum defined by inserting a Wilson line
does not depend on the shape of the path.

\begin{figure}[tb]
  \centering
  \includegraphics[width=4.5cm]{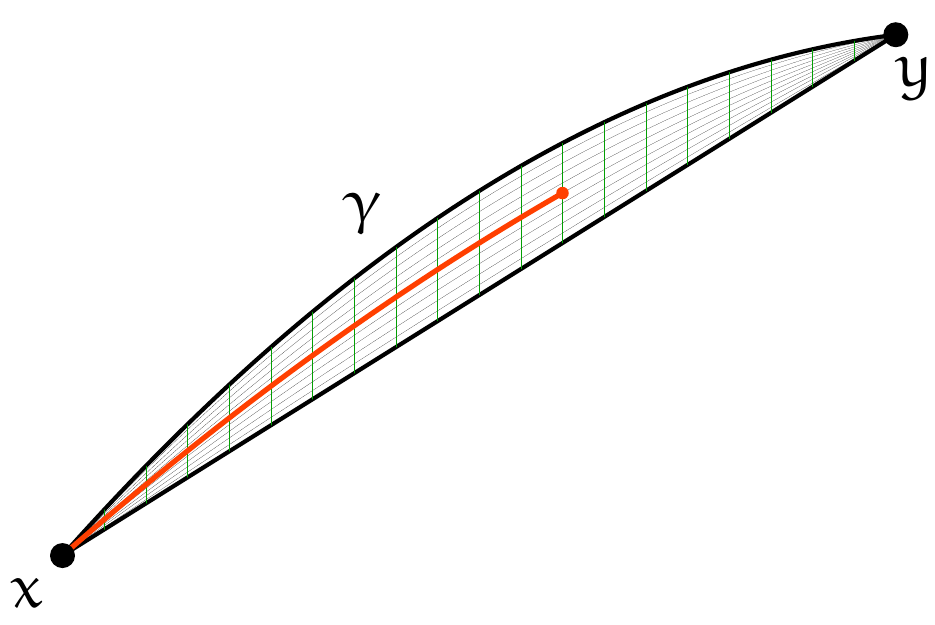}
  \hskip 1cm
  \includegraphics[width=4.5cm]{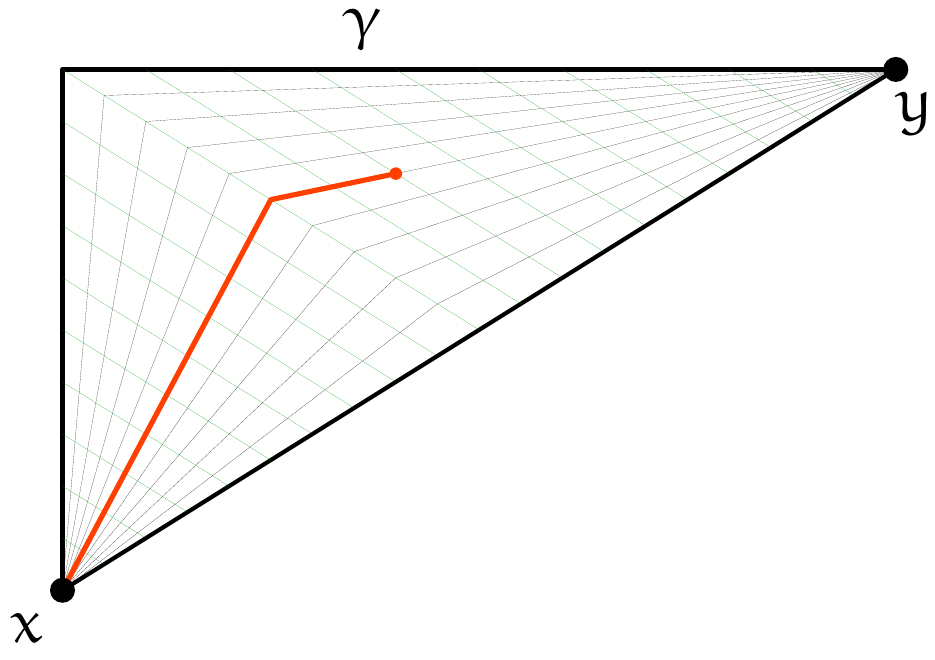}
  \hskip 1cm
  \includegraphics[width=2.31cm]{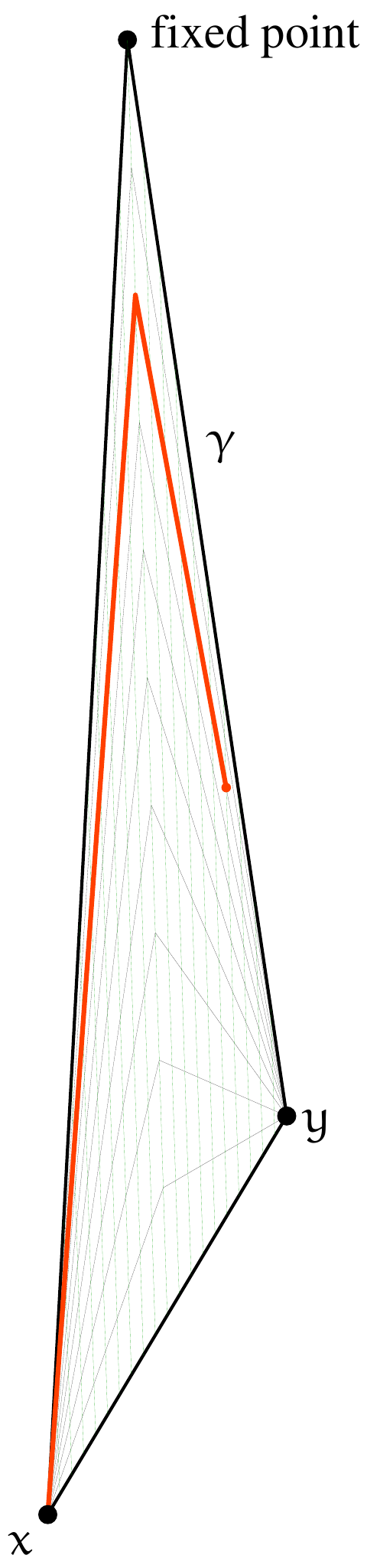}
  \caption{\label{fig:loops}Examples of paths that have different area
    and length scalings when $\y_\perp\to \x_\perp$. Left: smooth path
    with bounded curvature. Middle: path with unbounded curvature and
    a length that goes to zero linearly. Right: path with length that
    does not vanish when $\y_\perp\to \x_\perp$. In orange, we have
    represented a typical Wilson line $W_{\gamma(t,\cdot)}(0,s)$
    entering in the right hand side of eq.~(\ref{eq:nab-stokes}).}
\end{figure}

Let us now consider the general case. The Wilson line $W_\gamma$
defined on an arbitrary path may be related to the linear one by writing
\begin{align}
W_\gamma = \big(W_\gamma W_{_L}^{-1}\big)\,W_{_L}\, 
\end{align}
and the factor $W_\gamma W_{_L}^{-1}$ is a Wilson loop that can be
expressed in terms of the transverse magnetic field and the gauge
potentials $A^{1,2}$ thanks to the non-Abelian Stokes theorem. As we
have seen in the previous section, the important quantities are the
moments of the Fourier transform of the Wilson line, that are also the
Taylor coefficients of its short distance expansion. Therefore, one
way of assessing the effect of a curved Wilson line compared to a
straight one is to study how the loop $W_\gamma W_{_L}^{-1}$ behaves
when $\y_\perp\to \x_\perp$. Despite its complicated structure, the
non-Abelian Stokes formula is useful for estimating this behavior. Two
parameters control this limit. Firstly, because
$dtds \tfrac{\partial \gamma^\mu}{\partial t}\tfrac{\partial
  \gamma^\nu}{\partial s}$ measures the area of an elementary element
(e.g., the area of a little square in the figure \ref{fig:stokes}) of
the surface enclosed by the loop, there will be terms proportional to
the area times the magnetic field transverse to the loop.  Secondly,
because of the Wilson lines $W_{\gamma(t,\cdot)}(0,s)$, there will be
terms involving the gauge potential times the distance between
$\x_\perp$ and other points inside the loop. Because of this, we must
distinguish several possibilities for the shape of the path. Three of
them are illustrated in the figure \ref{fig:loops}, that we shall
study in turn hereafter.

\subsection{Paths with bounded curvature} The situation, shown on the
left panel of the figure \ref{fig:loops}, which is closest to a
straight line is a path whose curvature is bounded (i.e. whose radius
of curvature is always larger than some value $R$).  For such a path,
the area enclosed within the loop and the length of a typical Wilson
line $W_{\gamma(t,\cdot)}(0,s)$ scale as
\begin{align}
{\rm Area}\sim \frac{\delta^3}{R},\quad {\rm Length}\sim \delta,
\end{align}
where $\delta\equiv|\y_\perp-\x_\perp|$. When $\delta\to 0$, the Wilson
loop behaves as
\begin{align}
  W_\gamma W_{_L}^{-1}=1\oplus \frac{\delta^3 B}{R}\,\sigma^3\oplus
  \frac{\delta^4 BQ}{R}\,\sigma^3\sigma^{1,2}
  \oplus\cdots.
\label{eq:loop11}
\end{align}
In this formula, we have made explicit the fact that the longitudinal
magnetic field is proportional to the color generator $\sigma^3$.  At
the order $\delta^3$, we may replace by ${\bs 1}$ the Wilson line
$ W_{\gamma(t,\cdot)}(0,s)$ in the right hand side of
eq.~(\ref{eq:nab-stokes}), because its length goes to zero as
$\delta\to 0$. Deviations of this Wilson line from the identity
contribute only to higher orders in the separation ${\delta}$, e.g.,
the term in $\delta^4$ in eq.~(\ref{eq:loop11}), 
where $Q$ denotes generically the magnitude of $\Q_{1,2}$.  (Here, we
use the fact that the components of the gauge potential $A^i$ are
proportional to the color generators $\sigma^1$ or $\sigma^2$ in our
setup.) 
Likewise, the straight Wilson line behaves as
\begin{align}
W_{_L}=1\oplus \delta\; \big(Q\sigma^{1,2}\big)\oplus \delta^2\; \big(Q\sigma^{1,2}\big)^2\oplus\cdots.
\label{eq:loop12}
\end{align}
By multiplying eqs.~(\ref{eq:loop11}) and (\ref{eq:loop12}),
we get the following short distance behavior for a Wilson line with
bounded curvature,
\begin{align}
  W_\gamma
  &= \underbrace{1\oplus \delta\; \big(Q\sigma^{1,2}\big)\oplus \delta^2\; \big(Q\sigma^{1,2}\big)^2\oplus\cdots}_{W_{_L}}
    \notag\\
  &\qquad\oplus \frac{\delta^3 B}{R}\,\sigma^3
    \oplus \underline{\frac{\delta^4 B Q}{R}\sigma^3\sigma^{1,2}}\oplus \cdots
    \label{eq:loop13}
\end{align}
The first line in the right hand side is the result for a straight
Wilson line, studied more explicitly in the previous section. The
second line displays the structure of the first two terms involving
the magnetic flux (note that these terms all go to zero if
$R\to \infty$, i.e. if the curvature of the path under consideration
goes to zero). The first of these terms does not contribute, since 
it is proportional to $\sigma^3$ and only enters in the
following color trace,
\begin{align}
{\rm tr}\,\big(\sigma^3(1+\epsilon_1\sigma^1+\epsilon_2\sigma^2)\big)=0.
\end{align}
The second of these terms (underlined in eq.~(\ref{eq:loop13})) does
not suffer from this cancellation, since we have for instance
\begin{align}
  {\rm tr}\,\big(\sigma^3\sigma^1(1+\epsilon_1\sigma^1+\epsilon_2\sigma^2)\big)=
  2i\epsilon_2\not=0.
\end{align}
But note that this trace is proportional to $\epsilon_2$, and that
there is a sum on $\epsilon_2=\pm$ in the expression of the quark
spectrum. Therefore, to obtain a non-null contribution, another power
of $\epsilon_2$ must arise from the coefficient function
$G^{(n)}(\widehat\p_\perp,\k_\perp,\epsilon_1 Q_1,\epsilon_2
Q_2)$. The generic structure of $G^{(n)}$ reads
\begin{align}
G^{(n)}(\widehat\p_\perp,\k_\perp,\epsilon_1 Q_1,\epsilon_2
Q_2)
  =
  k_\perp^n \oplus \epsilon \underline{Q\;k_\perp^{n-1}} \oplus \cdots \oplus (\epsilon Q)^n.
\end{align}
But at the same time, recall that only terms with a degree in
$k_\perp$ equal to the order in $\delta$ can contribute, since we have
seen that the $p$-th moments of the Fourier transform of the Wilson
line are the Taylor coefficients of order $p$ in its short distance
expansion. We have underlined in the previous equation the term that
gives the leading high-$p_\perp$ behavior. Since the underlined term
in eq.~(\ref{eq:loop13}) is of order $\delta^4$, we must have $n=5$ to
obtain a matching term of order $k_\perp^4$, which means that the
magnetic flux starts to contribute only at the order
$p_\perp^{-5}$. Therefore, when the path defining the Wilson line has
a bounded curvature, the leading term in the high-$p_\perp$ quark
spectrum is universal, i.e. insensitive to variations of the shape of
the path. Schematically, the spectrum reads
\begin{align}
  \frac{dN}{dy_p d^2\p_\perp}
  \empile{=}\over{{\mbox{\scriptsize bounded}}\atop{\mbox{\scriptsize curvature}}}
  S_\perp \Bigg\{\underbrace{\frac{Q^4}{p_\perp^4}}_{\mbox{\scriptsize universal}}\oplus
  \underbrace{\frac{Q^2 B}{R\, p_\perp^5}}_{\mbox{\scriptsize path dependent}}\oplus\cdots \Bigg\}.
\label{eq:loop14}
\end{align}
Note that $B\sim Q^2$. But we prefer to keep explicitly the dependence
on the magnetic field, to stress the fact that the path dependent term
does not exist if the background field is purely electrical.

\subsection{Paths with unbounded curvature}
\label{sec:unbounded}
From the formula (\ref{eq:loop14}), we see that path dependent term
tends to grow when the radius of curvature decreases. As we shall now
show, if the curvature is unbounded, the path dependence is in fact
promoted at least one order earlier, at $p_\perp^{-4}$. In fact, there
are two main classes of paths with unbounded curvature: those whose
length goes to zero (linearly in $\delta$) when $\delta\to 0$, and
those whose length goes to a non-zero constant. Let us consider the
first class of paths, illustrated in the middle panel of figure
\ref{fig:loops}. In this case, we have
\begin{align}
{\rm Area}\sim {\delta^2},\quad {\rm Length}\sim \delta.
\end{align}
The reasoning is the same as in the previous subsection, but since the
area scales as $\delta^2$ instead of $\delta^3$, we need a term in
$\epsilon Q k_\perp^3$ (instead of $\epsilon Q k_\perp^4$) in the
final step. Such a term can be obtained in $G^{(4)}$, implying that
the path dependence now appears at the order $p_\perp^{-4}$ of the
high-$p_\perp$ spectrum,
\begin{align}
  \frac{dN}{dy_p d^2\p_\perp}
  \empile{=}\over{{\mbox{\scriptsize unbounded}}\atop{\mbox{\scriptsize curvature}}}
   S_\perp \Bigg\{\underbrace{\frac{Q^4}{p_\perp^4}}_{\mbox{\scriptsize universal}}\oplus
  \underbrace{\frac{Q^2 B}{ p_\perp^4}}_{\mbox{\scriptsize path dependent}}\oplus\cdots \Bigg\}.
\label{eq:loop24}
\end{align}
In Appendix~\ref{app:mag-unbounded}, we demonstrate that such a path-dependent $p_\perp^{-4}$ tail 
actually appears for specific choices of paths that have unbounded curvature.

\subsection{Paths going through a fixed point}
Consider now paths such as the one shown in the right panel of figure
\ref{fig:loops}, where $\gamma$ always goes through a fixed point. In
this case, we have
\begin{align}
{\rm Area}\sim \ell {\delta},\quad {\rm Length}\sim \ell,
\end{align}
where $\ell$ is the distance between $\x_\perp$ and this fixed
point. Since the length of the Wilson line $W_{\gamma(t,\cdot)}(0,s)$
does not go to zero, it contributes at a lower order than in the
previous two cases. Now, we can write 
\begin{align}
  W_\gamma W_{_L}^{-1}=1\oplus {\ell\delta B}\,\sigma^3\oplus \ell^2\delta BQ\sigma^3\sigma^{1,2}\oplus \cdots.
\label{eq:loop31}
\end{align}
The third term in the right hand side, that comes from the expansion
of $W_{\gamma(t,\cdot)}(0,s)$, has the same dependence on $\delta$ as
the second term. From this, we get
\begin{align}
  W_\gamma
  &= \underbrace{1\oplus \delta\; \big(Q\sigma^{1,2}\big)\oplus \delta^2\; \big(Q\sigma^{1,2}\big)^2\oplus\cdots}_{W_{_L}}
    \notag\\
  &\qquad\oplus {\delta\;\ell B}\,\sigma^3 
    \oplus \underline{\delta\;\ell^2 BQ\,\sigma^3\sigma^{1,2}}
    \oplus\cdots
\end{align}
Note that the color structure of the underlined term leads to a
non-zero trace. By the same reasoning as before, since this term is
linear in $\delta$, the adequate powers of $k_\perp$ can be found in
the coefficient function $G^{(2)}$. This means that for this type of
path, the path dependence would start at the order $p_\perp^{-2}$,
leading to an unphysically hard spectrum,
\begin{align}
  \frac{dN}{dy_p d^2\p_\perp}
  \empile{=}\over{{\mbox{\scriptsize fixed}}\atop{\mbox{\scriptsize point}}}
   S_\perp \Bigg\{\underbrace{\frac{\ell^2 Q^2 B}{p_\perp^2}}_{\mbox{\scriptsize path dependent}}\oplus
  \cdots \Bigg\}.
\label{eq:loop34}
\end{align}
\begin{figure}[htbp]
  \centering
  \includegraphics[width=10cm]{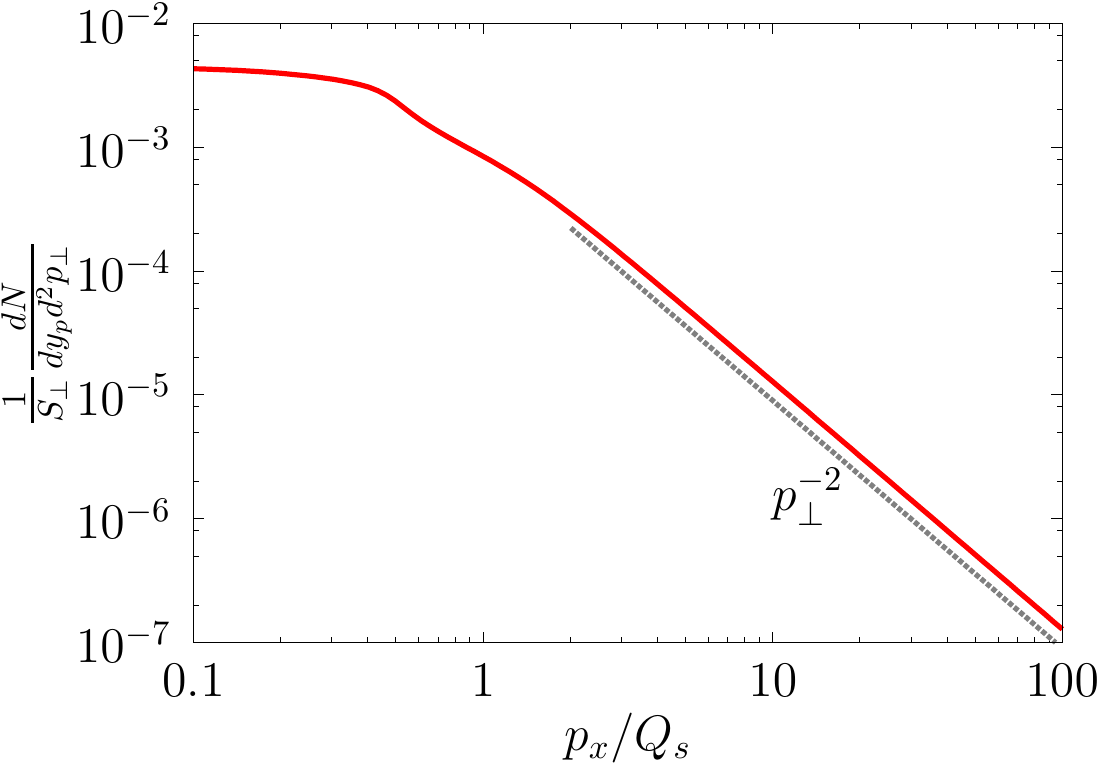}
  \caption{\label{fig:mag-fixed}Quark spectrum defined with a Wilson
    line that goes through a fixed point, in the purely magnetic case
    ($\Q_1\equiv\tfrac{1}{2}(0,Q_s), \Q_2\equiv \tfrac{1}{2}(Q_s,0),
    m/Q_s=0.1$). Solid red line: the spectrum \eqref{eq:spec_mag2} as a function of $p_x$ at
    $p_y=0$. Grey dotted line: power law $p_\perp^{-2}$.}
\end{figure}
The occurrence of such a term in $p_\perp^{-2}$ is illustrated in the
figure \ref{fig:mag-fixed}, where the points $\x_\perp$ and $\y_\perp$
are connected by the piecewise linear path $\gamma$ defined by
$\x_\perp\equiv(x^1,x^2)\to (0,x^2)\to(0,0)\to (0,y^2)\to \y_\perp\equiv(y^1,y^2)$.
See Appendix~\ref{app:mag-fixed} for details.

\section{Summary and conclusions}
\label{sec:concl}
In this paper, we have used a special configuration of $SU(2)$ color
currents in order to obtain an analytically tractable --at least up to
$\tau=0^+$, i.e. just after the collision has happened-- model of
two-nuclei collision in the CGC framework. Then, we evaluate the
inclusive quark spectrum at $\tau=0^+$ in the classical color field
obtained in this setup. Firstly, we have showed that without including
a Wilson line between the two spinors, this spectrum has an unphysical
tail in $p_\perp^{-2}$. Moreover, it does not vanish if we turn off
the color current of only one projectile. If a Wilson line defined on
a straight line connecting the two points is inserted, these
pathologies can be shown to disappear: the spectrum now decreases as
$p_\perp^{-4}$ and vanishes when any of the currents is turned off.

Then, we have studied the dependence of the spectrum on the shape of
the path, which influences the result when the background color field
has a magnetic component in the $z$ direction. In order to do this, we
first expressed the successive terms of the asymptotic expansion of
the spectrum in terms of the moments of the Fourier transform of the
Wilson line. These moments are then related to the short distance
expansion of the Wilson line in coordinate space. Using the
non-Abelian Stokes theorem, we showed that this short distance
behavior depends crucially on the shape of the path, notably on
whether it has a bounded curvature or is allowed to have kinks.

When the curvature is bounded, our power counting shows that the first
path dependent term is at most of order $p_\perp^{-5}$, implying that
the leading $p_\perp^{-4}$ term is shape independent. In this case,
the tail of the spectrum $p_\perp\gtrsim Q_s$ is universal even in the
presence of a non-pure gauge background field, but the softer part of
the quark spectrum may still suffer from an ambiguity related to the
choice of the path defining the Wilson line.

In contrast, if the curvature is not bounded, the path dependence is
promoted at least to order $p_\perp^{-4}$, or even to order
$p_\perp^{-2}$ if one considers a path that goes through a fixed
point. This means that with such a contour, the entire spectrum
--including the hard tail-- would be path dependent.  One should
presumably avoid using such singular paths, since it seems that the
unphysical features of the spectrum are a direct consequence of the
non-analyticity of the path.

Let us end by some more speculative words. Although our study was done
with a special configuration of the color sources of the projectiles,
the fact that the shape of the path has a strong influence on the
lowest order at which the path dependence arises is certainly robust.
We thus expect that, for completely general color backgrounds,
analytic curves are those that minimize the path dependence. Moreover,
although we have considered the quark spectrum at a proper time
$\tau=0^+$ in this paper in order to be able to perform some
analytical calculations, the same conclusions would hold for the
spectrum evaluated at a later time. The only change would be that,
thanks to the fact that the magnetic field decreases with time, the
path dependence should progressively decrease as $Q_s\tau$ increases.

\section*{Acknowledgements}
FG's work was supported by the Agence Nationale de la Recherche
through the project ANR-16-CE31-0019-01.

\appendix

\section{Asymptotic expansion of the function $G$}
\label{app:coeffs}
The first non-zero coefficient function, at the second order, reads
\begin{equation}
G^{(2)} (\widehat\p_\perp ,\k_\perp ,\Q_1 ,\Q_2)
= \frac{1}{2} ( \Q_1 +\Q_2 -\k_\perp )^2 -\frac{1}{3} \left[ \widehat\p_\perp \cdot ( \Q_1 +\Q_2 -\k_\perp )\right]^2 \, .
\end{equation}
Note that the result dependence of this coefficient on $\p_\perp$ is
only through the orientation of the transverse momentum, not its
magnitude.  At the third order, we can write
\begin{align}
  G^{(3)} (\widehat\p_\perp ,\k_\perp ,\Q_1 ,\Q_2)
  &\equiv
    G^{(3)}_0 (\widehat\p_\perp ,\k_\perp ,\Q_1 ,\Q_2)
    +
  G^{(3)}_1 (\widehat\p_\perp ,\k_\perp ,\Q_1 ,\Q_2)
    \notag\\
  &\quad
  +
  G^{(3)}_2 (\widehat\p_\perp ,\k_\perp ,\Q_1 ,\Q_2)
  +
  G^{(3)}_3 (\widehat\p_\perp ,\k_\perp ,\Q_1 ,\Q_2),
\end{align}
with
\begin{align}
&G^{(3)}_0 (\widehat\p_\perp ,\k_\perp , \Q_1 , \Q_2) \notag \\
&= -\frac{1}{6} \big[ 6  (\Q_1 \cdot \Q_2) ( \widehat\p_\perp \cdot \Q_1 +\widehat\p_\perp \cdot \Q_2) \notag\\
  &\hspace{30pt}
    + (\widehat\p_\perp \cdot \Q_1)(5\Q_1^2 -\Q_2^2) + (\widehat\p_\perp \cdot \Q_2)(5\Q_2^2 -\Q_1^2) \big] \notag \\
&\hspace{10pt}
+\frac{2}{3} \left[  (\widehat\p_\perp \cdot \Q_1)^3 + (\widehat\p_\perp \cdot \Q_2)^3
+ (\widehat\p_\perp \cdot \Q_1)^2 (\widehat\p_\perp \cdot \Q_2)
+ (\widehat\p_\perp \cdot \Q_1) (\widehat\p_\perp \cdot \Q_2)^2 \right],
\end{align}
\begin{align}
&G^{(3)}_1 (\widehat\p_\perp ,\k_\perp , \Q_1 , \Q_2) \notag \\
&= \frac{1}{6} \left[ (12   \Q_1 \cdot \Q_2 +5\Q_1^2 +5\Q_2^2 ) (\widehat\p_\perp \cdot \k_\perp ) \right. \notag \\
&\hspace{20pt} \left. 
+(10 \widehat\p_\perp \cdot \Q_1 +4 \widehat\p_\perp \cdot \Q_2 ) ( \Q_1 \cdot \k_\perp ) 
+(4 \widehat\p_\perp \cdot \Q_1 +10 \widehat\p_\perp \cdot \Q_2 ) ( \Q_2 \cdot \k_\perp )
\right] \notag \\
&\hspace{10pt}
-\frac{2}{3} \left[ 3 (\widehat\p_\perp \cdot \Q_1)^2 +3 (\widehat\p_\perp \cdot \Q_2)^2
+4  (\widehat\p_\perp \cdot \Q_1)(\widehat\p_\perp \cdot \Q_2) \right] (\widehat\p_\perp \cdot \k_\perp ),
\end{align}
\begin{align}
&G^{(3)}_2 (\widehat\p_\perp ,\k_\perp , \Q_1 , \Q_2)\notag \\
&= -\frac{1}{6} \left[ 5\left(  \widehat\p_\perp \cdot \Q_1 + \widehat\p_\perp \cdot \Q_2 \right) \k_\perp^2 +10(\widehat\p_\perp \cdot \k_\perp ) \left(  \Q_1 \cdot \k_\perp + \Q_2 \cdot \k_\perp \right) 
\right] \notag \\
&\hspace{10pt}
+2 \left(  \widehat\p_\perp \cdot \Q_1 + \widehat\p_\perp \cdot \Q_2 \right) (\widehat\p_\perp \cdot \k_\perp )^2,
\end{align}
and
\begin{align}
G^{(3)}_3 (\widehat\p_\perp ,\k_\perp , \Q_1 , \Q_2)
= \frac{5}{6} (\widehat\p_\perp \cdot \k_\perp ) \k_\perp^2 
-\frac{2}{3} (\widehat\p_\perp \cdot \k_\perp )^3.
\end{align}
The subscripts $0,1,2,3$ indicate the degree in $\k_\perp$ of the
corresponding terms. This organization is useful since it corresponds
to the successive coefficients of the Wilson line at small separations
$\x_\perp-\y_\perp$.

\section{Quark spectrum in a purely electric background} \label{app:elespec}
When the background field is purely electric, the quark spectrum \eqref{eq:spectrum-W} can be computed explicitly without relying on the asymptotic expansion in $1/p_\perp$. A purely electric background is realized if the two vectors $\Q_1$ and $\Q_2$ are parallel. In such a background, the Wilson line $W_\gamma (x,y)$ is independent of a path $\gamma$ and only a function of $\x_\perp -\y_\perp$ as 
\begin{align}
W_\gamma (\x_\perp ,\y_\perp )
&= \cos \left[ \bQbar \cdot (\x_\perp -\y_\perp ) \right] -i \frac{Q_1 \sigma^1 +Q_2 \sigma^2}{\sqrt{Q_1^2 +Q_2^2}} \sin \left[ \bQbar \cdot (\x_\perp -\y_\perp ) \right] .
\end{align}
Here, we denote $Q_{1,2}\equiv|\Q_{1,2}|$ and $\bQbar \equiv\sqrt{Q_1^2+Q_2^2}\, \Q_1 /Q_1$. The Fourier transform \eqref{eq:FTW} of this Wilson line is
\begin{align}
\wt{W}_\gamma (\k_\perp )
&= \frac{1}{2} S_\perp \sum_{\lambda=\pm} \left( 1+\lambda \frac{Q_1 \sigma^1 +Q_2 \sigma^2}{\sqrt{Q_1^2 +Q_2^2}} \right) (2\pi)^2 \delta (\k_\perp -\lambda \bQbar ) .
\end{align}
By plugging this expression into eq.~\eqref{eq:spectrum-W}, we find the quark spectrum in the purely electric background as
\begin{align}
\frac{dN}{dy_p d^2\p_\perp}
&= \frac{S_\perp}{(2\pi)^4} \sum_{\epsilon_1,\epsilon_2,\lambda=\pm} \left[ 1
-\left( 1+\lambda \frac{\epsilon_1 Q_1 +\epsilon_2 Q_2}{\sqrt{Q_1^2 +Q_2^2}} \right) 
G(\p_\perp ,\lambda \bQbar ,\epsilon_1 \Q_1, \epsilon_2 \Q_2 ) \right] .
\end{align}
One can easily confirm that this spectrum vanishes when one of $\Q_1$ or $\Q_2$ is zero.
In agreement with the general result \eqref{eq:4th-exp}, the high-$p_\perp$ expansion of this spectrum starts with the order $p_\perp^{-4}$,
\begin{equation}
\frac{dN}{dy_p d^2\p_\perp} = \frac{S_\perp}{60\pi^4 p_\perp^4}
\left[ 15 Q_1^2 Q_2^2 -8(\widehat{\p}_\perp \cdot \Q_1)^2 (\widehat{\p}_\perp \cdot \Q_2)^2 \right]
+\mathcal{O} (p_\perp^{-6}) .
\end{equation}

\section{Quark spectrum in a purely magnetic background}
\label{app:magspec}
When the vectors $\Q_1$ and $\Q_2$ are not parallel, nonzero chromo-magnetic fields are generated after the collision. In the presence of a magnetic field, the Wilson line has a path-dependence and it is not easy in general to compute the Fourier transform of the Wilson line analytically. 
In this appendix, we consider a purely magnetic field configuration that is realized by
\begin{equation}
\Q_1 = (Q_1 ,0) \, , \hspace{10pt}
\Q_2 = (0 ,Q_2) ,
\end{equation} 
and derive explicit forms of the quark spectra for specific paths that allow analytic calculations.

\subsection{Paths with unbounded curvature} \label{app:mag-unbounded}
First, we take a one-parameter family of paths which are piecewise linear,
\begin{equation*}
  \gamma_\alpha~:\quad \x_\perp \equiv(x^1 ,x^2)
  \to (\alpha x^1 +(1-\alpha)y^1, x^2)
  \to (\alpha x^1 +(1-\alpha)y^1, y^2)
  \to \y_\perp\equiv(y^1, y^2) , 
\end{equation*}
with a varying parameter $\alpha \in [0,1]$, as shown here:
\begin{equation*}
\includegraphics[width=8cm]{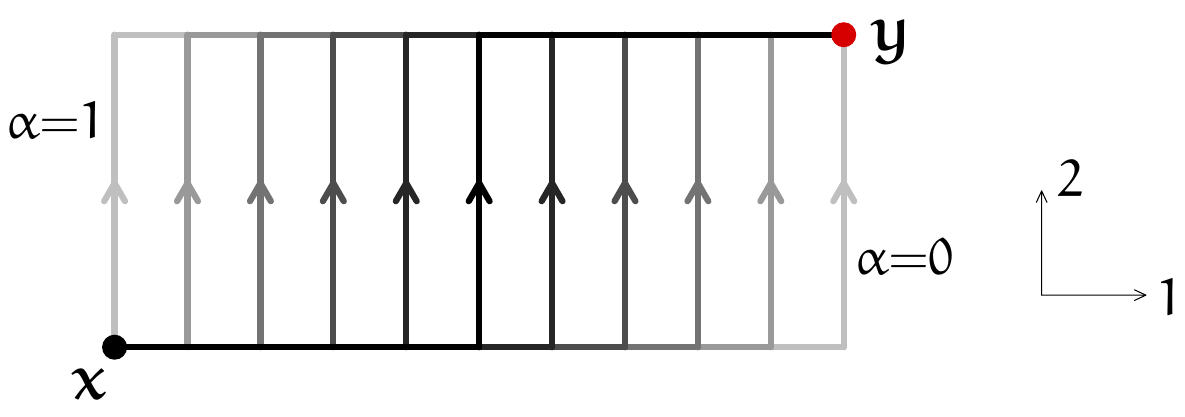}
\end{equation*}
The paths in this family
have an unbounded curvature, but a length that goes to zero as
$\y_\perp\to\x_\perp$.  Therefore, according to the discussion
presented in the section~\ref{sec:unbounded}, we expect the that quark
spectrum contains an $\alpha$-dependent term of order $p_\perp^{-4}$.
Using the shorthand notations $\Delta x^1\equiv x^1-y^1$ and
$\Delta x^2\equiv x^2-y^2$, the Wilson line along this contour reads
\begin{align}
W_{\gamma_\alpha} (\x_\perp ,\y_\perp ) 
&= e^{-i(1-\alpha) Q_1 \sigma^1 \Delta x^1} e^{-iQ_2 \sigma^2 \Delta x^2} e^{-i\alpha Q_1 \sigma^1 \Delta x^1} \notag \\
&= \frac{1}{4} \sum_{\lambda_1,\lambda_2=\pm} \left[ (1+\lambda_1 \sigma^1)e^{-i\lambda_1 Q_1 \Delta x^1 -i\lambda_2 Q_2 \Delta x^2} \right. \notag \\
&\hspace{60pt} \left. 
+\lambda_2 (\sigma^2 +i\lambda_1 \sigma^3 ) e^{-i\lambda_1 (1-2\alpha) Q_1 \Delta x^1 -i\lambda_2 Q_2 \Delta x^2} \right] .
\end{align}
Its Fourier transform is
\begin{align}
\wt{W}_{\gamma_\alpha} (\k_\perp )
&= \frac{S_\perp}{4} \sum_{\lambda_1,\lambda_2=\pm} \left[ (1+\lambda_1 \sigma^1) (2\pi)^2 \delta (\k_\perp -\lambda_1 \Q_1-\lambda_2 \Q_2) \right. \notag \\
&\hspace{60pt} \left. 
+\lambda_2 (\sigma^2 +i\lambda_1 \sigma^3 ) (2\pi)^2 \delta (\k_\perp -\lambda_1(1-2\alpha) \Q_1-\lambda_2 \Q_2)  \right] .
\end{align}
For this Wilson line, the quark spectrum \eqref{eq:spectrum-W} becomes
\begin{align}
\frac{dN}{dy_p d^2 \p_\perp}
&= \frac{S_\perp}{32\pi^4} \sum_{\epsilon_1,\epsilon_2,\lambda_1,\lambda_2=\pm} \big[
1-(1+\lambda_1 \epsilon_1) G(\p_\perp , \lambda_1 \Q_1+\lambda_2 \Q_2, \epsilon_1 \Q_1, \epsilon_2 \Q_2 ) \notag \\
&\hspace{90pt}
-\lambda_2 \epsilon_2 G(\p_\perp , \lambda_1 (1-2\alpha) \Q_1+\lambda_2 \Q_2, \epsilon_1 \Q_1, \epsilon_2 \Q_2 ) \big] .
\label{eq:spec_mag1}
\end{align}
The high-$p_\perp$ asymptotic form of this spectrum is
\begin{align}
\frac{dN}{dy_p d^2 \p_\perp}
&= \frac{S_\perp Q_1^2 Q_2^2}{60\pi^4 p_\perp^4} \left( 25 -8\widehat{p}_x^2 \widehat{p}_y^2 \right)
-\frac{S_\perp Q_1^2 Q_2^2}{15\pi^4 p_\perp^4} (2\alpha-1)^2 \left(15-32\widehat{p}_x^2 \widehat{p}_y^2 \right) +\mathcal{O} (p_\perp^{-6}) .
\label{eq:spec_mag1exp}
\end{align}
This result is perfectly consistent with eq.~\eqref{eq:loop24}. The
first term of the right hand side corresponds to the universal term
given in eq.~\eqref{eq:4th-exp}, and the second term depends on the
parameter $\alpha$ that characterizes the shape of the contour. Note
that when $\alpha=1/2$, the path dependent term vanishes and the
asymptotic form of the spectrum coincides with that computed with a
straight Wilson line, given in eq.~\eqref{eq:4th-exp}.
\begin{figure}[htbp]
 \centering
  \includegraphics[width=10cm]{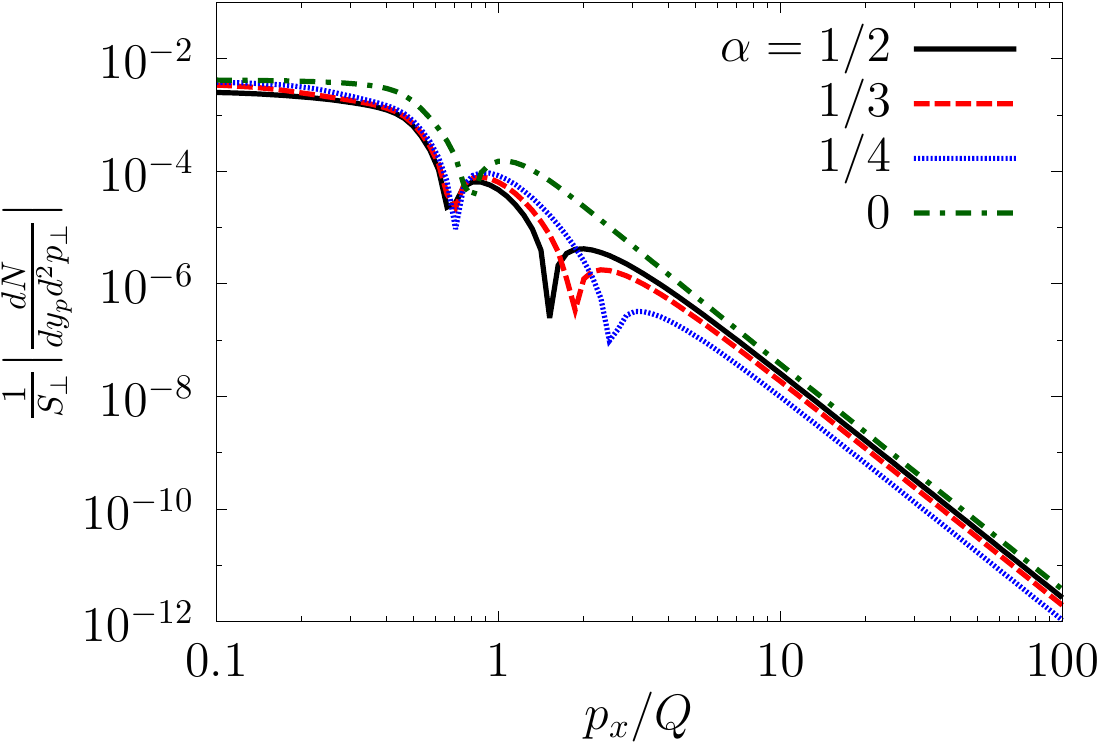} 
  \caption{\label{fig:specB1}Comparison of the spectrum in the pure
    magnetic background \eqref{eq:spec_mag1} for different paths
    parameterized by $\alpha=0$, 1/4, 1/3, 1/2. The parameters are set
    to $Q_1 =Q_2 =\tfrac{1}{2} Q_s$ and $m/Q_s=0.1$. We show the absolute
    value of the spectrum since it is not positive definite over the
    entire momentum range with this type of Wilson line.}
\end{figure}
In the figure~\ref{fig:specB1}, the spectrum \eqref{eq:spec_mag1} is
plotted as a function of $p_x$ for several values of $\alpha$. Since
the spectrum depends on $\alpha$ only through $(2\alpha-1)^2$, it is
sufficient to consider $\alpha \in [0,1/2]$.
We can confirm that the $p_\perp^{-4}$ tails depend on the path parameter $\alpha$.

\subsection{Paths going through a fixed point} \label{app:mag-fixed}
Next, as an explicit example of path that goes through a fix point, we
consider a piecewise linear path defined by
\begin{equation*}
\gamma~:\quad 
\x_\perp\equiv(x^1,x^2)\to (0,x^2)\to(0,0)\to (0,y^2)\to \y_\perp\equiv(y^1,y^2) ,
\end{equation*}
as shown here:
\begin{equation*}
\includegraphics[width=6cm]{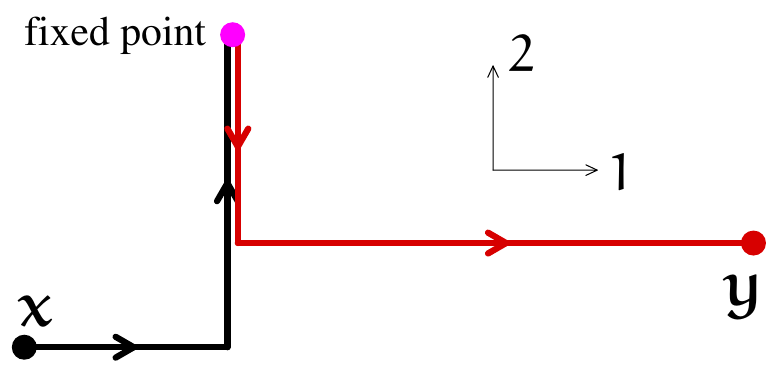}
\end{equation*}
In a pure magnetic background, the Wilson line on this path is
\begin{equation}
W_\gamma (\x_\perp ,\y_\perp )
= e^{-iQ_1 \sigma^1 x^1} e^{-iQ_2 \sigma^2(x^2-y^2)} e^{iQ_1 \sigma^1 y^1} .
\end{equation}
Its Fourier transform is
\begin{equation}
\wt{W}_\gamma (\k_\perp ) 
= \frac{S_\perp}{4} \sum_{\lambda_1,\lambda_2=\pm} (1+\lambda_1 \sigma^1) (2\pi)^2 \delta (\k_\perp -\lambda_1 \Q_1-\lambda_2 \Q_2)
\end{equation}
for $Q_1\neq 0$, and the quark spectrum reads
\begin{align}
\frac{dN}{dy_p d^2 \p_\perp}
&= \frac{S_\perp}{32\pi^4} \sum_{\epsilon_1,\epsilon_2,\lambda_1,\lambda_2=\pm} \big[
1-(1+\lambda_1 \epsilon_1) G(\p_\perp , \lambda_1 \Q_1+\lambda_2 \Q_2, \epsilon_1 \Q_1, \epsilon_2 \Q_2 )  \big] .
\label{eq:spec_mag2}
\end{align}
This spectrum has an unphysical $p_\perp^{-2}$ tail at high $p_\perp$ as shown in the figure \ref{fig:mag-fixed}.


\end{document}